\documentclass[amsmath,amssymb,aps,
prb,
twocolumn,
floatfix,
longbibliography,
floatfix
]{revtex4-2}

\usepackage{graphicx,dcolumn,bm,color,multirow,hyperref}

\begin{document}

\title{{\it Ab initio} study of lattice dynamics of group IV semiconductors using pseudohybrid functionals for extended Hubbard interactions}

\author{Wooil Yang}
\affiliation{Department of Physics, 
Pohang University of Science and Technology, 
Pohang 37673, Korea
}
\author{Seung-Hoon Jhi}
\email{jhish@postech.ac.kr}
\affiliation{Department of Physics, 
Pohang University of Science and Technology, 
Pohang 37673, Korea
}

\author{Sang-Hoon Lee}
\affiliation{
Korea Institute for Advanced Study, 
Seoul 02455, Korea
}
\author{Young-Woo Son}
\email{hand@kias.re.kr}
\affiliation{
Korea Institute for Advanced Study, 
Seoul 02455, Korea
}

\date{\today}

\begin{abstract}
We study the lattice dynamics of group IV semiconductors using fully {\it ab-initio} extended Hubbard functional. The onsite and intersite Hubbard interactions are determined self-consistently with recently developed pseudohybrid functionals and included in force calculations. We analyze the Pulay forces by the choice of atomic orbital projectors and the force contribution of the onsite and intersite Hubbard terms. The phonon dispersions, Gr\"uneisen parameters, and lattice thermal conductivities of diamond, silicon, and germanium, which are most-representative covalent-bonding semiconductors, are calculated and compared with the results using local, semilocal, and hybrid functionals. The extended Hubbard functional produces increased phonon velocities and lifetimes, and thus lattice thermal conductivities compared to local and semilocal functionals, agreeing with experiments very well. Considering that our computational demand is comparable to simple local functionals, this work thus suggests a way to perform high-throughput electronic and structural calculations with a higher accuracy.
\end{abstract}

\maketitle

\section{Introduction\label{sec:intro}}
Density functional theory (DFT)~\cite{1,2} has been very successful in prediction of various physical properties of real materials~\cite{Jones2015RMP}. Popular practical implementations of DFT treat the electron correlation at non-interacting one electron level within the local density approximation (LDA)~\cite{2} or the generalized gradient approximation (GGA)~\cite{3}. A common but most serious problem that may arise from the approximations of electron interaction in LDA and GGA is the errors from improper treatment of the electron self-interaction known as self-interaction errors (SIE)~\cite{4,5}. Not only for strong-correlated systems but also for conventional covalent-bonding materials, the SIE leads to incorrect description of the ground systems. Underestimation or overestimation of the lattice constant or the energy band-gap are well-known syndromes of these functionals. Such incorrect calculation of the lattice calculation also leads to the errors in elastic properties and phonon spectrums. 
	
In order to alleviate the shortcomings associated with the SIE, various modified functionals such as meta-GGA functionals~\cite{6,7,8,9,10,11} or hybrid functional~\cite{12,13,14} were developed. These approaches still contain empirical parameters to handle the long-range screening and fail to describe strongly correlated localized systems and low dimensional systems~\cite{Jain2011PRL}. Another level of approximation to treat the many-body electron correlation is to include the dynamic screening in Coulomb interaction as in the $GW$ method~\cite{16,17,18} and DMFT~\cite{15}. These methods yet require significant computational resources compared to standard DFT, impractical for massive calculations of many-atom systems
and for extensive high-throughput calculations.
	
Since SIE is related to the issue of electron localization, a common remedy is to compensate the delocalization error particularly for $d$ or $f$ electron systems by introducing the local Coulomb repulsion or the on-site Hubbard $U$ interaction. The DFT$+U$ method corrects the issue of electron over-delocalization, capturing the essential physics in the Mott insulators and the magnetic ground states~\cite{Anisimov1991PRB,60,Dudarev1998PRB,19}. However, it fails to describe the conventional covalent-bonding semiconductors~\cite{22,21,20,Timrov2021PRB} since the hybridization of the extended orbitals between the pair of atoms is not handled well by the Hubbard $U$ term for the localized sites. The intersite Coulomb interaction between the localized orbitals, or the extended Hubbard $V$ term, should be considered on equal footing for proper description of the covalent bonding~\cite{22,21,20,Timrov2021PRB}. DFT with extended Hubbard functionals was shown to give the band gap of conventional semiconductors accurately in a similar level of $GW$ method with significantly reduced computational demand~\cite{22,21,20,Timrov2021PRB}.  

In typical DFT$+U$ methods, the on-site Hubbard $U$ was chosen empirically in a reasonable range to produce the standard structural or electronic properties such as the lattice constant and band gap matched to experimental measurement~\cite{Anisimov1991PRB,60,Dudarev1998PRB,19}. Recently, various methods to compute Hubbard parameters self-consistently have been proposed~\cite{23,24,25,26,27,28,29,30,31}. Among them, the pseudohybrid Hubbard density functional proposed by Agapito-Curtarolo-Buongiorno Nardelli (ACBN0) leads to a direct self-consistent determination of $U$ using Hartree-Fock (HF) formalism with moderate computational cost~\cite{31}. In a similar way, the intersite Hubbard $V$ term is implemented into the ACBN0 functional~\cite{21,20}. The extended ACBN0 functional determines self-consistently the Hubbard parameter $U$ and $V$ and calculates the band gaps of diverse semiconductors and insulators comparable to those from $GW$ methods in much less computational load than $GW$ calculations~\cite{21,20}. The self-consistent determination of the Hubbard terms enables accurate computation of lattice dynamics and structural relaxation beyond the one-particle level approximation of LDA or GGA. 

In this work, using the newly developed extended Hubbard functional, we carried out comparative study of the structural and electronic properties of group IV semiconductors diamond (C), silicon (Si), and germanium (Ge) to make point-by-point comparison with various exchange-correlation functionals. These semiconductors are the most representative covalent-bonding semiconductors, expected to exhibit the apparent correction from the inter-site Hubbard term $V$. We were particularly interested in the lattice dynamics of these semiconductors such as the phonon band structure, the mode Gr\"uneisen parameters, and the thermal conductivity as they are affected by the Hubbard term, $U$ and $V$. The computed static and dynamic lattice properties are in excellent agreement with experiment results, thus demonstrating effectiveness as well as accuracy of the newly developed extended Hubbard functional in studying electronic and structural properties of solids simultaneously. 

This paper is organized as follows. We first briefly review our formalism of the pseudohybrid functionals for 
intersite Hubbard interactions and associated forces in Sec.~\ref{sec:formalism}. 
The detailed computational parameters for calculations of electron energy bands as well as
static and dynamic lattice properties are listed in Sec.~\ref{sec:comp}.
Then, using the new method described in Sec.~\ref{sec:formalism}, 
we present our computational results of electronic, structural and phonon properties of group IV semiconductors in Sec.~\ref{sec:results}.
Finally, we conclude in Sec.~\ref{sec:conclusion}.

\section{Formalism for extended Hubbard energy functionals and forces \label{sec:formalism}}

Here we briefly review the DFT method with $U$ and $V$ interactions~\cite{22,21,20,Timrov2021PRB}
and a recently developed intersite Hubbard pseudohybrid functional~\cite{21,20}. Then, the forces originating from $U$ and $V$ terms are discussed. 
We first start by considering the total energy formula with the Hubbard energy functional, 
\begin{equation}
E_\textrm{tot}=E_\textrm{DFT}+E_\textrm{Hub}.
\label{Eq:tot}
\end{equation}
In Eq.~\ref{Eq:tot}, $E_\textrm{DFT}$ can be any local or semilocal density functional and the rotational invariant Hubbard functional, $E_\textrm{Hub}$~\cite{Dudarev1998PRB,22} can be written with 
the $U_\textrm{eff}\equiv U-J$ and $V$ with double counting corrections,
\begin{eqnarray}
E_\textrm{Hub} &=&\frac{1}{2}\sum_I \sum_{i,j,\sigma} U_{\textrm{eff}}^I (\delta_{ij}-n^{II\sigma}_{ij})n^{II\sigma}_{ji}\nonumber\\
&&-\frac{1}{2}\sum_{\{I,J\}}\sum_{i,j,\sigma}V^{IJ}n^{IJ\sigma}_{ij}n^{JI\sigma}_{ji},
\label{Eq:Hub}
\end{eqnarray}
where the general occupation matrix is written as
\begin{eqnarray}
n^{IJ\sigma}_{ij}&\equiv& n^{I,n,l,J,n',l',\sigma}_{ij}\nonumber \\
&=& \sum_{m{\bf k}}f_{m\bf k}
\langle \psi_{m{\bf k}}^\sigma |\phi^{I,n,l}_i \rangle \langle \phi^{J,n',l'}_j |\psi_{m{\bf k}}^\sigma\rangle \nonumber\\
&\equiv& \sum_{m{\bf k}}f_{m\bf k}
\langle \psi_{m{\bf k}}^\sigma |\phi^{I}_i \rangle \langle \phi^{J}_j |\psi_{m{\bf k}}^\sigma\rangle.
\label{Eq:GO}
\end{eqnarray}
Here $f_{m\bf k}$ is the Fermi-Dirac function of 
the Bloch state $|\psi_{m\bf k}^\sigma\rangle$ 
of the $m$-th band at a momentum ${\bf k}$.
In Eq.~\ref{Eq:Hub}, $\{I,J\}$ denotes a pair of different atoms within a cutoff distance. 
The principle ($n$), azimuthal ($l$), angular ($i$), and spin ($\sigma$) quantum numbers  
of the $I$-th atom are implicitly written in the last line of Eq.~\ref{Eq:GO}, that will be used hereafter. 
The L{\"o}wdin orthonormalized atomic orbital (LOAO) of $\phi^I_i$~\cite{32,33} is used
as a projector.

To obtain pseudohybrid functionals for Hubbard interactions, we follow an ansatz by Mosey {\it et al.}~\cite{29,30} and ACBN0 functional~\cite{31}
that leads to `renormalized' occupation number ($N^{IJ\sigma}_{\psi_{m{\bf k}}}$) and density matrix ($P^{IJ\sigma}_{ij}$) for the pair of different atoms $I$ and $J$ such as
\begin{eqnarray}
  N^{IJ\sigma}_{\psi_{m{\bf k}}} & = &\sum_{I,i} \langle \psi_{m{\bf k}}^\sigma |\phi^{I}_i \rangle \langle \phi^{I}_i |\psi_{m{\bf k}}^\sigma\rangle \nonumber\\
  & & +\sum_{J,j}
   \langle \psi_{m{\bf k}}^\sigma |\phi^{J}_j \rangle \langle \phi^{J}_j |\psi_{m{\bf k}}^\sigma\rangle,\label{Eq:ansatz}\\
   P^{IJ\sigma}_{ij} &=& \sum_{m{\bf k}}f_{n\bf k} N^{IJ\sigma}_{\psi_{m{\bf k}}}
\langle \psi_{m{\bf k}}^\sigma |\phi^I_i \rangle \langle \phi^J_j |\psi_{m{\bf k}}^\sigma\rangle.
\label{Eq:RON}
\end{eqnarray}
For the pair of same atoms or for $U$ calculation, the above expression reduces to $N^{I\sigma}_{\psi_{m{\bf k}}} = \sum_{I,i} \langle \psi_{m{\bf k}}^\sigma |\phi^{I}_i \rangle \langle \phi^{I}_i |\psi_{m{\bf k}}^\sigma\rangle$.

The HF energy can be expressed with Eqs.~\ref{Eq:ansatz} and~\ref{Eq:RON} and the bare Coulomb repulsion between electrons belong to $i$ and $k$ orbitals
of atom $I$ and $j$ and $l$ orbitals of atom $J$,
$
(ik|jl)\equiv\int d{\bf r}_1 d{\bf r}_2 
\frac{\phi^{I*}_i ({\bf r}_1)\phi^I_k({\bf r}_1)\phi^{J*}_j ({\bf r}_2)\phi^J_l ({\bf r}_2)}{|{\bf r}_1-{\bf r}_2|}.
\label{Eq:ERI}
$
By inspecting equivalence between unrestricted HF formulation and Dudarev form of extended Hubbard interactions, 
we can obtain following functional forms for $U^I$, $J^I$~\cite{31} and $V^{IJ}$~\cite{21,20} as follows,
\begin{eqnarray}
U^I &=& \sum_{ijkl}{\mathcal U}^I_{ijkl} (ij|kl),
~J^I = \sum_{ijkl}{\mathcal J}^I_{ijkl} (ik|jl),
\label{Eq:UJ}\\
V^{IJ}&=& \sum_{ij}{\mathcal V}^{IJ}_{ij}(ii|jj)\label{Eq:V}.
\end{eqnarray}
It is immediately noticeable that Hubbard functionals in Eqs.~\ref{Eq:UJ} and~\ref{Eq:V} are weighted Coulomb interactions between electrons belong to orbitals of the pair atoms. The weight factors can be expressed with Eqs.~\ref{Eq:GO},~\ref{Eq:ansatz} and~\ref{Eq:RON} such that 
$
{\mathcal U}^I_{ijkl} =
\frac{1}{\mathcal N^I}
\sum_{\sigma,\sigma'} P^{II\sigma}_{ij} P^{II\sigma'}_{kl},
$
$
{\mathcal J}^I_{ijkl} =
\frac{1}{\mathcal N^J}
\sum_\sigma P^{II\sigma}_{ij} P^{II\sigma}_{kl}
$
and
$
{\mathcal V}^{IJ}_{ij}=\frac{1}{2\mathcal N^{IJ}}
\sum_{\sigma,\sigma'} 
[
P^{II\sigma}_{ii}P^{JJ\sigma'}_{jj}-\delta_{\sigma \sigma'}P^{IJ\sigma}_{ij}P^{JI\sigma'}_{ji}
]
$
where normalization factors are
${\mathcal N}^I = \sum_{i\neq j} \sum_{\sigma} n^{II\sigma}_{ii}n^{II\sigma}_{jj}
+\sum_{ij}\sum_{\sigma}n^{II\sigma}_{ii}n^{II-\sigma}_{jj}$,
${\mathcal N^J}=
\sum_{i\neq j} \sum_\sigma
n^{II\sigma}_{ii}n^{II\sigma}_{jj}$,
and 
${\mathcal N^{IJ}}=
\sum_{\sigma,\sigma'}\sum_{ij} 
[
n^{II\sigma}_{ii} n^{JJ\sigma'}_{jj}-\delta_{\sigma \sigma'}n^{IJ\sigma}_{ij}n^{JI\sigma'}_{ji}
]$.
We note that the weight factors could be regarded as the position dependent mixing parameters of HF interactions reflecting local changes of Coulomb interactions as well as nonlocal variation for intersite screenings.

Forces from extended Hubbard interactions on the $K$-th atom can be obtained using derivative of $E_\textrm{Hub}$ in Eq.~\ref{Eq:Hub}
with respect to its displacement of ${\bf R}_K$~\cite{20,39}. 
Using the chain rule,
\begin{eqnarray}
\frac{\partial E_\textrm{Hub}}{\partial {\bf R}_K}
&=&\sum_I \sum_{i,j,\sigma}
\frac{\partial E_\textrm{Hub}}
{\partial n^{II\sigma}_{ij}}
\frac{\partial n^{II\sigma}_{ij}}
{\partial {\bf R}_K} 
 +\sum_I \frac{\partial E_\textrm{Hub}}
{\partial U^I_\textrm{eff}}
\frac{\partial U^I_\textrm{eff}}
{\partial {\bf R}_K}\nonumber\\
& &+\sum_{\{I,J\}} \frac{\partial E_\textrm{Hub}}
{\partial V^{IJ}}
\frac{\partial  V^{IJ}}
{\partial {\bf R}_K}.
\end{eqnarray}
After some algebra, each contribution to the total Hubbard force, ${\bf F}_\textrm{Hub} \equiv 
\partial E_\textrm{Hub}/\partial {\bf R}_K
={\bf F}^N+{\bf F}^U+{\bf F}^V$ can be written as,
\begin{eqnarray}
{\bf F}^N
 &\equiv&\sum_I \sum_{i,j,\sigma} U_{\textrm{eff}}^I \left(\frac{\delta_{ij}}{2}-n^{II\sigma}_{ij}\right)
 \frac{\partial n^{II\sigma}_{ji}}{\partial {\bf R}_K}\nonumber
 \\
&&-\sum_{\{I,J\}}\sum_{i,j,\sigma}V^{IJ}n^{IJ\sigma}_{ij}
\frac{\partial n^{JI\sigma}_{ji}}{\partial {\bf R}_K}
\label{Eq:Pulay} 
\\
{\bf F}^U
&\equiv&\frac{1}{2}\sum_I \sum_{i,j,\sigma}\frac{\partial U_\textrm{eff}^I}{\partial {\bf R}_K} (\delta_{ij}-n^{II\sigma}_{ij})n^{II\sigma}_{ji},
\label{Eq:UForce}
\\
{\bf F}^V
&\equiv&-\frac{1}{2}\sum_{\{I,J\}}\sum_{i,j,\sigma}\frac{\partial V^{IJ}}{\partial {\bf R}_K} n^{IJ\sigma}_{ij}n^{JI\sigma}_{ji}.
\label{Eq:VForce}
\end{eqnarray}
In calculating
$\partial n^{II\sigma}_{ji}/\partial {\bf R}_K$ in Eq.~\ref{Eq:Pulay}, the main contribution comes from a derivative of LOAO, 
$\partial \phi^{I,n,l}_i /\partial {\bf R}_K$ so that ${\bf F}^N$ can be regarded as the Pulay force~\cite{39}.
Unlike nonorthorgonalized atomic orbital (NAO) projectors, $\partial n^{II\sigma}_{ji}/\partial {\bf R}_K$ in Eq.~\ref{Eq:Pulay} is not zero in case of $I\neq J\neq K$~\cite{39} so do $\partial U^I_\textrm{eff}/\partial {\bf R}_K$ in Eq.~\ref{Eq:UForce} and $\partial V^{IJ}/\partial {\bf R}_K$
in Eq.~\ref{Eq:VForce}. Difficulty in evaluating the derivative of LOAO~\cite{20,39} has been overcome by Timrov {\it et al.}~\cite{39} recently so that the direct evaluation of ${\bf F}^N$ is now feasible. 

Typically, ${\bf F}^U$ and ${\bf F}^V$ are quite small and have been neglected so far~\cite{20,39,89} while for some metal oxide molecules, ${\bf F}^U$ is not negligible~\cite{kulik2011JCP}. Since we can compute the onsite and intersite Hubbard interactions self-consistently, we can check orders of magnitudes of forces from Eq.~\ref{Eq:UForce} and Eq.~\ref{Eq:VForce} directly. For the semiconducting materials here, the forces from the variation of directional bonding is the most significant so that we can simplify Eq.~\ref{Eq:UForce} and Eq.~\ref{Eq:VForce} as 
\begin{eqnarray}
{\bf F}^U
&\simeq&\frac{1}{2} \sum_{i,j,\sigma}\frac{\partial U_\textrm{eff}^K}{\partial {\bf R}_K} (\delta_{ij}-n^{KK\sigma}_{ij})n^{KK\sigma}_{ji},
\label{Eq:UForceApprox}
\\
{\bf F}^V
&\simeq&-\frac{1}{2}\sum_{\{J\}}\sum_{i,j,\sigma}\frac{\partial V^{KJ}}{\partial {\bf R}_K} n^{KJ\sigma}_{ij}n^{JK\sigma}_{ji},
\label{Eq:VForceApprox}
\end{eqnarray}
where $\{J\}$ in Eq.~\ref{Eq:VForceApprox} indicates a sum of contribution of $J$-th atom whose distance with respect to $K$-th atom is within a given cut-off distance.

\section{Computational details\label{sec:comp}}

All DFT calculations are performed with {\sc Quantum Espresso}~\cite{34} and norm-conserving pseudopotentials (NC-PP) from Pseudo Dojo library~\cite{35}. In the case of LDA, PBEsol, and the extend Hubbard functionals, the Brillouin zone sampling was done on a $k$-point grid of $15\times15\times15$ mesh to calculate the equilibrium lattice parameter and bulk modulus, and on a $k$-point grid of $21\times21\times21$ mesh for electronic structure calculations. For Heyd-Scuseria-Ernzerhof (HSE) hybrid functional calculations, we used $7\times7\times7$ grids for both calculations. The volume dependence of the static lattice energy was fitted to Vinet's equation of state~\cite{36}. The band structure in HSE calculations was obtained via Wannier interpolation~\cite{37} using {\sc Wannier90}~\cite{38}. The Hubbard $U$ and $V$ parameters were calculated self-consistently using the in-house version of  {\sc Quantum Espresso}~\cite{21} with the on-site $U$ for $s$ orbitals set to be zero.
The cut-off energy is set to be 100 Ry and the self-consistency threshold for the total energy and Hubbard interactions is $10^{-8}$ Ry. The cut-off distance for intersite $V$ is set to include the nearest neighbors, which is enough for convergence of total energy of the group IV semiconductors~\cite{21}.

The self-consistent $U$ and $V$ terms are expressed in terms of the localized-orbital projectors, and the site-dependent Pulay forces naturally arise. The choice of the atomic orbital projects is thus a critical step in practical implementation for calculating $U$ and $V$ terms and the forces. Here we employed the method implemented by Timrov {\it et al.}~\cite{39} that calculates the Pulay force and stress using orthogonalized atomic wave functions as projectors. For addressing lattice dynamics, the Pulay force and the derivatives of the Hubbard $U$ and $V$ terms should be calculated correctly in addition to the standard DFT forces. We analyzed the contribution of each force and the errors from inaccurate estimate of the occupation number in the interstitial regions. It is found that the Hubbard forces from the derivative of $U$ and $V$ is very small and thus neglected in the force calculations as discussed in Appendix~\ref{append:pulay}. 

The harmonic and cubic anharmonic interatomic force constants (IFCs) of phonons were calculated using the frozen-phonon method~\cite{40,41} in a supercell of 64 atoms with a $k$-point grid of $5\times5\times5$ and $3\times3\times3$ mesh, respectively. For HSE calculations, the $k$-grid was reduced to $2\times2\times2$ due to heavy computational cost. The thermal conductivity was calculated using the phonon Boltzmann transport equation (BTE)~\cite{41,42,43,44,45,46,47,48}. We employed the relaxation time approximation (RTA)~\cite{49} to calculate the phonon lifetime for Si and Ge since the Umklapp scattering is relatively strong around the room temperature~\cite{42,50}. However, for C, the normal process dominates to Umklapp process~\cite{43,47,48}, and its lattice thermal conductivity within RTA is severely underestimated compared with the full converged solution~\cite{43}. The lattice thermal conductivity of C was calculated with the direct solution of the linearized phonon BTE~\cite{51,52}. 

\section{Computational Results\label{sec:results}}

\begin{table}[b]
\caption{
Calculated $U$ and $V$ for $s$ and $p$ orbitals (in eV). Up, the on-site Hubbard term for $p$ orbital and, $V_{ss}$, $V_{sp}$, and $V_{pp}$, the intersite Hubbard term between $s$-$s$, $s$-$p$, and $p$-$p$ orbitals of the nearest atoms, respectively.
}
\centering
\begin{ruledtabular} 
\begin{tabular}{ccccc}
     & $U_p$  &  $V_{ss}$ & $V_{sp}$  & $V_{pp}$           \\
      \hline
C   &  5.91    & 0.82         & 1.09          & 2.92  \\
Si  &  3.51    & 0.88         & 0.72         & 1.86    \\
Ge & 3.33       & 1.03     & 0.70 & 1.76
\end{tabular}
\end{ruledtabular}
\label{table:UV}
\end{table}

Calculated $U$ and $V$ values of C, Si, and Ge at the experimental lattice constants are presented on Table~\ref{table:UV}. The $U$ and $V$ of Si are almost the same with a previous study~\cite{21}. Our calculated values are in good agreement with the ones in the literature. We note that the intersite $V$ term is about 25$\sim$50\% of the onsite $U$ term in these covalent semiconductors and thus should be correctly determined for accurate description of physical properties. Before proceeding to the lattice dynamics, we calculated the lattice constants, the bulk modulus, and the electronic band structures with various exchange-correlation for point-by-point comparison.

\subsection{Structural parameters\label{subsec:structure}}

\begin{table}[t]
\caption{Optimized theoretical lattice constants in \AA~calculated with LDA, PBE, PBEsol, HSE, and extended Hubbard functional ($U+V$). Experimental values are measured at ambient conditions. 
}
\centering
\begin{ruledtabular} 
\begin{tabular}{ccccccc}
     & LDA & PBE & PBEsol & HSE & $U+V$ & Exp.        \\
      \hline
C   & 3.537  & 3.572 & 3.558  & 3.558 & 3.562 & 3.567\footnote{Reference~\onlinecite{Skinner1957}}\\
Si  & 5.394  & 5.469 & 5.431  & 5.434 & 5.434 & 5.431\footnote{Reference~\onlinecite{Becker1982}} \\
Ge & 5.621  & 5.764 & 5.676 & 5.629 & 5.662 & 5.658\footnote{Reference~\onlinecite{62}}
\end{tabular}
\end{ruledtabular}
\label{table:lattice}
\end{table}

\begin{table}[b]
\caption{Optimized bulk modulus in GPa calculated with LDA, PBE, PBEsol, HSE, and extended Hubbard functional ($U+V$).  Experimental values are measured at ambient conditions. 
}
\centering
\begin{ruledtabular} 
\begin{tabular}{ccccccc}
     & LDA & PBE & PBEsol & HSE & $U+V$ & Exp.        \\
      \hline
C   & 465 & 432 & 449 & 464 & 450 & 442\footnote{Reference~\onlinecite{Grimsditch1975}}\\
Si  & 96 & 88 & 93 & 99 & 95 & 96$\sim$ 99.4\footnote{Reference~\onlinecite{Senoo1976, Singh1977}} \\
Ge & 72 & 59 & 67 & 74 & 72  & 75$\sim$ 75.8\footnote{Reference~\onlinecite{Bruner1961, McSkimin1963}}
\end{tabular}
\end{ruledtabular}
\label{table:bulkmodulus}
\end{table}

Calculated lattice constant and bulk modulus of group IV elements calculated with various functionals are presented in Tables~\ref{table:lattice} and~\ref{table:bulkmodulus} together with experiment for comparison. Each functional using NC-PP gives the similar result with the previous DFT calculations using various pseudopotentials~\cite{53,54,55,56,57}; LDA underestimates the lattice constants and thus gives hardened bulk modulus compared with experiment while PBE overestimates the lattice constants and consequently gives soft bulk moduli. PBEsol and HSE give improved lattice constant and bulk modulus. Compared to these one-electron-based functionals, DFT with the onsite $U$ and intersite Hubbard $V$ terms gives very accurate lattice constants and bulk moduli within 0.2 and 3 \% of errors overall compared with measurement. The extended Hubbard functional describes the covalent bonding character properly not only at static level but also at dynamic level as we discuss below.

\subsection{Electronic band structures\label{subsec:band}}

\begin{figure*}
	\begin{center}
		\includegraphics[width=1.7\columnwidth]{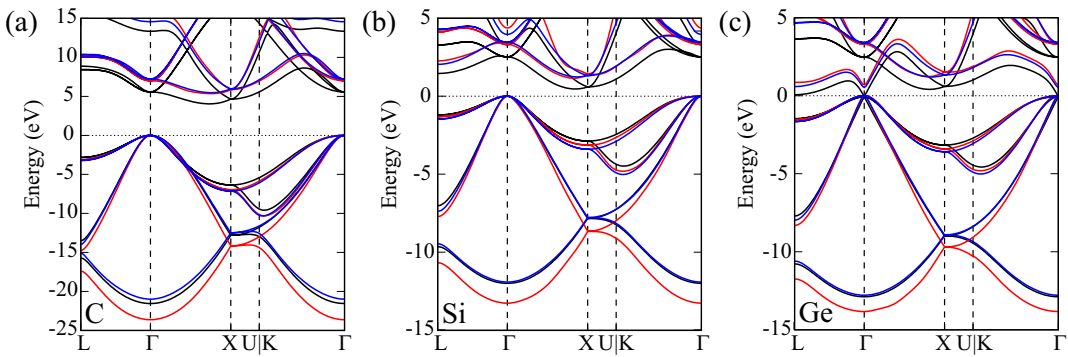}
	\end{center}
	\caption{Calculated electronic band structures of C, Si, and Ge along the symmetry lines in the Brillouin zone using PBEsol (black), HSE (red), and the extend Hubbard functional (blue). The top of the valence band is set to 0 eV.
	}
	\label{fig:band}
\end{figure*}

Our calculated electronic band structures of the group IV semiconductors are shown in Fig.~\ref{fig:band}, and the fundamental energy gaps are summarized in Table~\ref{table:bandgap}. The range of our calculated band gaps are comparable with experimental data. The onsite Hubbard $U$ does not improve the energy band gaps of semiconductors~\cite{22,21,20}. It is clear that the intersite Hubbard $V$ term increases the band gap of the group IV semiconductors consistent with previous reports~\cite{20,21,22}. For proper description of the covalent $sp^3$ hybridization, the intersite interaction should be included. We note that, for Ge, the band gap is calculated without considering the spin-orbit interaction and the minimum gap is direct at $\Gamma$.

The highest valence band in the extended Hubbard functional is shifted slightly lower than those of PBEsol, HSE, and $GW$ approximation~\cite{58}. Overall the valence bands in the extended Hubbard functional exhibit similar dispersions as in PBEsol and also in $GW$ calculations (Si and Ge)~\cite{58}. We observe wide band-width in the valence bands and a rigid shift of about 1 eV of the lowest-lying valence band in HSE results due to the strong coupling between the valence $p$ and the conduction $s$ band~\cite{53,59}. On the other hand, the onsite $U$ of the $s$ orbital is zero and no meaningful shift of the lowest lying $s$ band from the PBEsol band is observed in the extended Hubbard functional. 

The conduction bands in the extended Hubbard functional show a rigid shift to higher energies compared to those in PBEsol. The lowest conduction band in the extended Hubbard functional is similar to that in HSE but slightly lower in energy than that in $GW$ method~\cite{58}. The extend Hubbard functionals may be regarded as a static approximation to the GW approximation~\cite{21,20,60} expected to produce comparable accuracy to the $GW$ method.

\begin{table*}
\caption{Calculated and measured band gaps (Exp) in eV unit. $E_g^d$, direct band gap at $\Gamma$ point; $E_g^i$, the indirect band gap. For Ge, the spin-orbit interaction is not considered in the calculation and the minimum gap is direct at $\Gamma$.
}
\centering
\begin{ruledtabular}
\begin{tabular}{ccccccccc}
 &\multicolumn{2}{c}{PBEsol} &\multicolumn{2}{c}{HSE}&\multicolumn{2}{c}{$U+V$}&\multicolumn{2}{c}{Exp.}\\
\cline{2-9}
 & $E_g^i$ & $E_g^d$ & $E_g^i$ & $E_g^d$ & $E_g^i$ & $E_g^d$ & $E_g^i$ & $E_g^d$ \\
 \hline
C & 4.04 & 5.56 & 5.36 & 7.00 & 5.47 & 7.22 & 5.48\footnote{Reference~\onlinecite{C.D.Clark1964}} & 7.3\footnote{Reference~\onlinecite{Roberts1967}} \\
Si & 0.46 & 2.51 & 1.15 & 3.32 & 1.23 & 3.46 & 1.12\footnote{Reference~\onlinecite{McLean1960}} & 3.40\footnote{Reference~\onlinecite{Daunois1978}} \\
Ge& 0.13 & 0.01 & 0.85 & 0.70 & 0.59 & 0.54 & 0.66\footnote{Reference~\onlinecite{Macfarlane1957}} &0.80\footnote{Reference~\onlinecite{Camassel1975}}
\end{tabular}
\end{ruledtabular}
\label{table:bandgap}
\end{table*}

\subsection{Phonon band structures and Gr\"uneisen parameters}

We studied the lattice dynamics of group IV semiconductors to demonstrate how the intersite Hubbard $V$ term describes the covalent bonding properly. All calculations were performed at the experimental lattice constant of C (3.5670 \AA), Si (5.4310 \AA) both at room temperature~\cite{Skinner1957, Becker1982}, and Ge (5.6524 \AA) at 80 $K$ ~\cite{62}. Normally, the optimized lattice constant by a specific functional is used for the lattice dynamics and electronics structures. However, apparent good agreement with experiment sometimes comes from the cancellation, for instance in LDA case, of the underestimation of the lattice constants and the vibrational frequencies to some extent~\cite{53}. Here we chose to confirm the correction by the intersite Hubbard $V$ term in the extended Hubbard functional compared to LDA, PBE, PBEsol, and HSE under the same condition. Using the finite-difference method, we calculated the IFCs up to the second order for the phonon band structures and the third order for the Gr\"uneisen parameters. In the extended Hubbard functional, the Pulay forces arise inevitably from the local atomic orbital basis, and we employed the recently developed method by Timrov {\it et al}.~\cite{39} to handle the orthogonal atomic orbitals. In the extended Hubbard functional, implementation of the self-consistent calculation of $U$ and $V$ has naturally been done with atom-centered orthogonal orbitals, but the force calculation can be complicated (see Appendix~\ref{append:orbital} on the analysis of the latent errors from inaccurate estimate of the occupation number in the overlap regions on the Pulay forces).

\begin{figure*}
	\begin{center}
		\includegraphics[width=1.5\columnwidth]{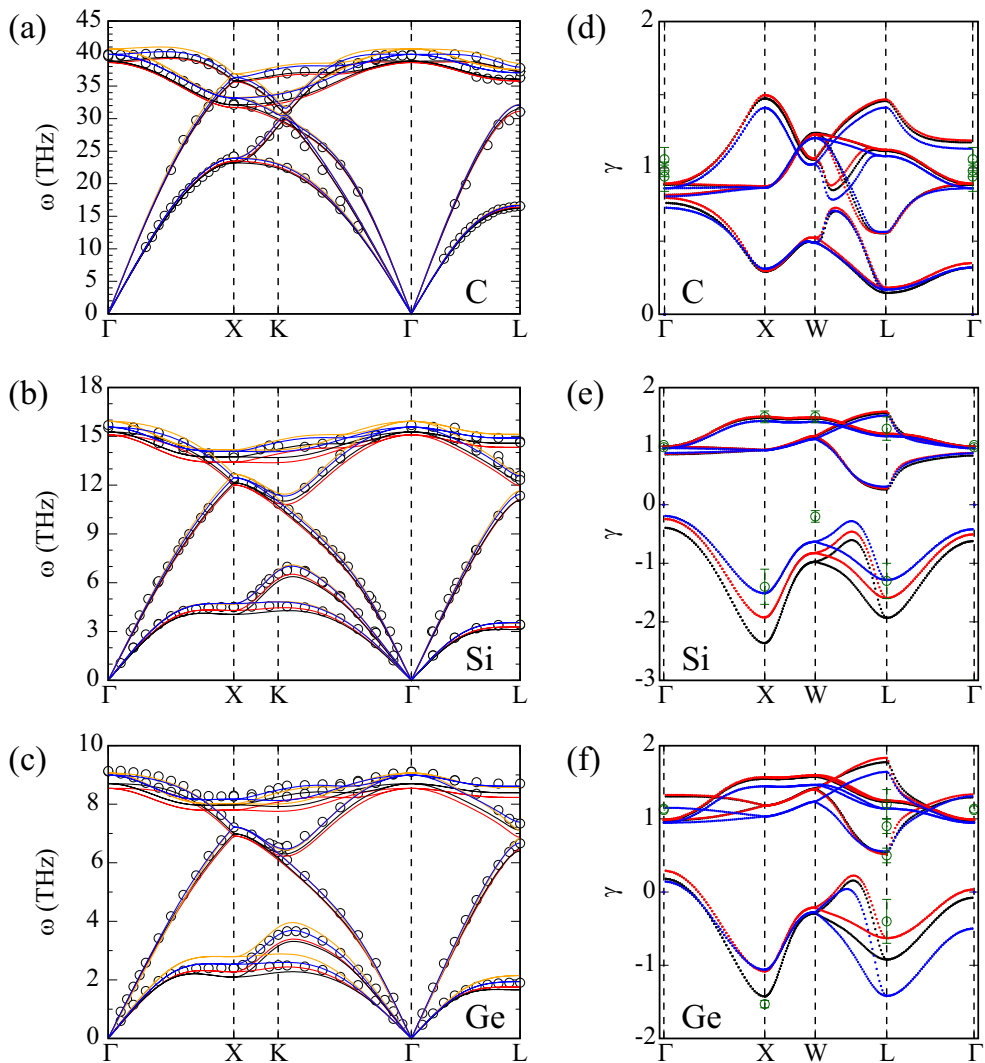}
	\end{center}
	\caption{Calculated phonon band structures (left column) and the mode Grüneisen parameters ($\gamma$) (right column). From top to bottom, C, Si, and Ge. The phonon bands were obtained at the experimental lattice constant with the exchange-correlation functional of LDA (red), PBEsol (black), HSE (orange), and the extended Hubbard functional (blue). Open circles denote the experimental values; C~\cite{75,79,80}, Si~\cite{76,77,81}, and Ge~\cite{78}. The mode Grüneisen parameters were obtained using cubic IFCs with LDA (red), PBEsol (black), and the extended Hubbard functional (blue). Open circles in green denote the measured values of C~\cite{82,83,84}, Si~\cite{85,86}, and Ge~\cite{85,87,88}.}
	\label{fig:phonon}
\end{figure*}

Figures~\ref{fig:phonon} (a)-(c) shows calculated phonon band structures at the experimental lattice constants. Calculated phonon bands with the extend Hubbard functional are in very good agreement with experiment compared to the bands by other functionals. Overall, the local and semilocal functionals underestimate the phonon frequencies and nonlocal HSE overestimates them. As such, the phonon bands by the extended Hubbard functional are located between those by LDA/PBEsol and HSE. Slight overestimation of phonon frequencies by HSE is due to enhancement of interatomic interaction between neighboring atoms by the inclusion of nonlocal exact exchange compared to PBEsol~\cite{53}.

\begin{table*}
\caption{Our calculated phonon frequencies at high symmetry points obtained at the experimental lattice constants. Measurements are from Refs.~\onlinecite{75,76,77,78}. 
}
\centering
\begin{ruledtabular}
\begin{tabular}{cccccccccc}
  &	&$\Gamma_{LO/TO}$ & $X_{TA}$ & $X_{LA/LO}$ & $X_{TO}$ & $L_{TA}$ & $L_{LA}$ &	$L_\text{LO}$ & $L_\text{TO}$ \\
  \hline
\multirow{5}{*}{C} &LDA&	38.6	&23.4&35.7&31.7&16.4&31.1&	37.3	&35.8\\
&PBEsol&	38.8&	23.2&	35.8&	32.1&	16.2&	31.5&	37.3&	36.1\\
&HSE&	40.7&	24.0&	36.9&	33.2&	16.7&	32.2&	38.5&	37.6\\
&$U+V$&	39.9&	23.9&	36.3&	33.1&	16.5&	32.1&	37.9&	37.1\\
&Exp.\footnote{Reference~\onlinecite{75}} &	40.3&	24.2&	36.1&	32.6&	16.4&	31.0&	37.2&	36.3\\
\hline
\multirow{5}{*}{Si}								
&LDA&	15.1&	4.3&	12.0&	13.4&	3.3&	11.1&	12.0&	14.3\\
&PBEsol&	15.3&	4.1&	12.1&	13.7&	3.1&	11.1&	12.3&	14.6\\
&HSE&	15.9&	4.7&	12.7&	14.2&	3.5&	11.7&	12.7&	15.1\\
&$U+V$&	15.6&	4.7&	12.5&	14.1&	3.5&	11.4&	12.6&	14.9\\
&Exp.\footnote{References~\onlinecite{76,77}}&	15.5&	4.5&	12.3&	13.9&	3.4&	11.4&	12.6&	14.7\\
\hline
\multirow{5}{*}{Ge}							
&LDA&	8.54&	2.28&	6.91&	7.80&	1.75&	6.44&	6.87&	8.23\\
&PBEsol&	8.70&	2.10&	6.99&	7.98&	1.66&	6.40&	7.11&	8.39\\
&HSE&	9.06&	2.79&	7.21&	8.02&	2.15&	6.84&	7.09&	8.58\\
&$U+V$&	8.99&	2.53&	7.22&	8.17&	1.94&	6.67&	7.39&	8.61\\
&Exp.\footnote{Reference~\onlinecite{78}}&	9.11&	2.40&	7.22&	8.27&	1.89&	6.66&	7.34& 8.69
\end{tabular}
\end{ruledtabular}
\label{table:phonon}
\end{table*}

The phonon frequencies at high symmetry points are shown in Table~\ref{table:phonon} for detailed comparison. We note that our calculated phonon frequencies of diamond using the extended Hubbard functional of 39.9, 36.3, and 37.9 THz at $\Gamma$, $X$, and $L$ point, respectively, are comparable to the harmonic phonon frequencies obtained with the variational quantum Monte Carlo method of 40.7, 36.5, and 38.0 THz~\cite{63}.

We next calculated the mode Gr\"uneisen parameters of C, Si, and Ge from cubic IFCs to see how well the extended Hubbard functional describes the response of the interatomic forces in covalent bonding systems to volume changes. Calculated mode Gr\"uneisen parameters together with available experimental data are plotted in Figs.~\ref{fig:phonon}(d)-(f). All the functionals here give negative Gr\"uneisen parameters in the transverse acoustic modes for Si and Ge, which explains their negative thermal expansion~\cite{64,65}. We observe quite a difference between the functionals in the transverse acoustic branches of Si and Ge. Apparently, the variation in the Gr\"uneisen parameters between the functionals reflects the difference in the phonon dispersions of the low-frequency transverse acoustic modes. Compared to experiment, the extended Hubbard functional reproduces very accurately the small Gr\"uneisen parameters of Si, which is related with its weak anharmonicity. While all functionals give agreeable results with the experimental values at the $\Gamma$, $X$, $W$, and $L$ points in the transverse acoustic modes, the extended Hubbard functional produces more accurate results than LDA or PBEsol functionals. As the transverse acoustic modes are characterized by the non-central covalent bonding~\cite{64}, inclusion of the $V$ term in Hubbard functionals captures such features of the covalent bonding appropriately.

\subsection{Lattice thermal conductivity and phonon lifetime\label{subsec:thermalcond}}

\begin{figure}
	\begin{center}
		\includegraphics[width=1\columnwidth]{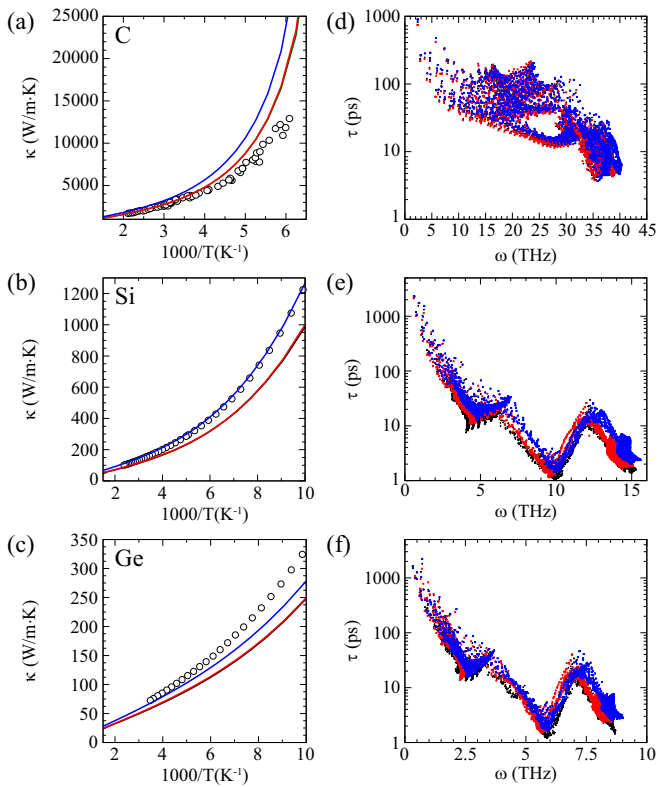}
	\end{center}
	\caption{Calculated lattice thermal conductivity ($\kappa$) (left column) as a function of temperature ($T$) and phonon life-time ($\tau$) (right column) as a function of frequency ($\omega$) of diamond, Si, and Ge, from top to bottom, respectively. PBEsol, black; LDA, red; the extended Hubbard functional, blue. The PBEsol and LDA give almost identical results. Experimental values of lattice thermal conductivity (C~\cite{66,67}, Si~\cite{68}, and Ge~\cite{69,70}) are plotted in open circles.
	}
	\label{fig:thermal}
\end{figure}

Finally we investigated the thermal transport using the extended Hubbard functional to compare it with other functionals. Figure~\ref{fig:thermal} shows calculated lattice thermal conductivity and phonon lifetime together with experimental results. Because of weak Umklapp scattering relative to the normal scattering in C, the thermal conductivity obtained from BTE within RTA is severely underestimated~\cite{43,47,48}, and instead we used the direct solution of the linearized phonon BTE~\cite{51,52}.

Our LDA and PBEsol calculations are almost identical, and produce similar or slightly smaller values for C at low temperature but underestimated for Si and Ge compared to experiment~\cite{66,67,68,69,70}. This result is due to our choice of the experimental lattice constants for phonon calculations to analyze the functional dependence at the same footing. The phonon group velocity and thus the thermal conductivity are very sensitive to the lattice constants. It was shown that the increase in the lattice constant by 1\% in Si results in a decrease of the lattice thermal conductivity by ~4\%~\cite{71}. The apparent agreement of the LDA calculation with experiment in previous studies~\cite{43,72,73} is likely attributed to error cancellation of underestimated lattice constants. Also, while calculated lattice thermal conductivity within RTA is close to the full BTE solutions when the Umklapp scattering is dominant, it is still smaller by as much as ~5\% than the full BTE solutions for Si and Ge~\cite{74}. The detailed values of our calculations are off from the experimental results, but the tendency of the thermal conductivity matches well with experiment. 

Calculated thermal conductivity with the extended Hubbard functional is slightly larger than those by LDA and PBEsol over all temperature ranges considered (left column of Fig. 4). The thermal conductivity is determined basically by the phonon group velocity and lifetime. Our calculation shows that this behavior in the thermal conductivity mostly comes from the difference in the phonon lifetime. In the entire frequency range, we observe that the extended Hubbard functional produces larger phonon lifetime than LDA or PBEsol (right column of Fig. 4). Our calculated mode Gr\"uneisen parameters with the extended Hubbard functional are smaller than those with LDA or PBEsol, and this result indicates the weaker phonon-phonon scattering and thus longer lifetimes for phonons depicted in the extended Hubbard functional. The phonon group velocity also contributes to the thermal conductivity as well since the phonon bands calculated with the extended Hubbard functional have steeper dispersion in the acoustic branches than those by LDA or PBEsol. Our calculations support that the intersite interactions in the covalent-bonding semiconductors are accurately described by the extended Hubbard functional. The lattice constant, the phonon dispersion, lifetime, and the lattice thermal conductivity are properly produced by the extended Hubbard functional compared to experiment.

\section{Conclusion\label{sec:conclusion}}

We studied the lattice dynamics of group IV elements using fully {\it ab initio} extended Hubbard functional. Calculated electronic band structures show that the intersite Hubbard $V$ term is essential in describing covalent-bonding orbital hybridization. Equilibrium lattice parameters, bulk modulus, and energy gaps calculated with the extended Hubbard functional match with experiment better than those with other functionals. 

Dynamical properties of group IV elements are also correctly described when the intersite $V$ term in the extended Hubbard functional is considered. Our calculated phonon frequencies with the extended Hubbard functional show better agreement with experiment than those with local, semilocal or hybrid functionals. Calculated mode Gr\"uneisen parameters with the extended Hubbard functional show relatively large difference from those with other functionals, especially in the acoustic branches of Si and Ge, where the values are negative. We expect that the extended Hubbard functional provides accurate description of the thermal expansion of Si and Ge. We found that the phonon group velocity and life-time calculated with the extended Hubbard functional are larger than those with other functionals, and so is the lattice thermal conductivity. 

We note that the derivate of $U$ and $V$ terms contributes little to the total forces and is not included in our force calculations. Rather, the Pulay forces from the derivatives of the occupation number are found to be crucial. For full consistency, the derivative of $U$ and $V$ with respect to atomic displacements may be necessary, in particular, for more correlated systems than conventional semiconductors.

\section*{Acknowledgement}
Y.-W.S. thanks Bo Gyu Jang and Shi Liu for fruitful discussion. 
S.-H.J. was supported by the National Research Foundation of Korea (NRF) (Grant No. 2018R1A5A6075964) funded by the Korea government (MSIT). 
Y.-W.S. was supported by NRF of Korea (Grant No. 2017R1A5A1014862, SRC program: vdWMRC center) and KIAS individual Grant No. (CG031509). 
Supercomputing resources including technical supports were provided by Supercomputing Center, Korea Institute of Science and Technology Information (Contract No. KSC-2020-CRE-0173)
and the Center for Advanced Computation of Korea Institute for Advanced Study.

\appendix

\begin{table*}
\caption{The Hubbard forces (in Ry/Bohr radius) for the displacement along [111] direction. $R_K$, Si-Si displacement (in Bohr radius); $F^U + F^V$, the derivative of $U$ and $V$. The Pulay forces are calculated with orthogonal (LOAO) and nonorthogonal atomic (NAO) basis with elongated $(+)$ and contractive $(-)$ displacement. Total forces are also shown to be compared with the Hubbard forces.
}
\centering
\begin{ruledtabular}
\begin{tabular}{cccccccc}
 & & \multicolumn{4}{c}{LOAO} & \multicolumn{2}{c}{NAO}\\
 \cline{3-8}
 $R_K$ & $F^U + F^V$ & \multicolumn{2}{c}{$F^{N}$} & \multicolumn{2}{c}{$F_\textrm{tot}$} & \multicolumn{2}{c}{$F^{N}$}\\
 \cline{3-8}
 & & $(+)$ & $(-)$ & $(+)$ & $(-)$& $(+)$ & $(-)$ \\
 \hline
0.025&	0.000032&	$-0.000005$	&0.000003&	$-0.007029$	&0.007314&	$-0.000192$ &	0.000185\\
0.050	&0.000044&	$-0.000014$	&0.000002&	$-0.013778$	&0.014914	&$-0.000390$&	0.000361\\
0.075&	0.000064&	$-0.000026$	&$-0.000004$	&$-0.020249$&	0.022804&	$-0.000596$&	0.000528\\
0.100&	0.000101&	$-0.000042$&	$-0.000014$&	$-0.026447$	&0.030989&	$-0.000867$&	0.000684
\end{tabular}
\end{ruledtabular}
\label{table:force}
\end{table*}

\section{Comparison of Pulay forces and the derivative of U and V\label{append:pulay}}
We calculated the Hubbard force contribution from simplified ${\bf F}^U$ and ${\bf F}^V$ in Eq.~\ref{Eq:UForceApprox} and Eq.~\ref{Eq:VForceApprox} for Si. Table~\ref{table:force} shows calculated $F^U=|{\bf F}^U|$ and $F^V=|{\bf F}^V|$, and $F^N=|{\bf F}^N|$ using NAO and LOAO.
$R_K=|{\bf R}_K|$
The cutoff radius for evaluating the intersite Hubbard term $V$ is set to include only the nearest neighbor atoms in the ground configuration. $F^U$ and $F^V$ are calculated using the finite difference method with small displacement of $R_K=|{\bf R}_K|$ along $[111]$ direction. $F^U + F^V$ are similar in magnitude with $F^N$ calculated with LOAO, and both are much smaller than total forces~$F_\textrm{tot}=|dE_\textrm{tot}/d{\bf R}_K|$.

\section{Atomic orbital basis in the calculation of Hubbard forces\label{append:orbital}}

\begin{figure}[t]
	\begin{center}
		\includegraphics[width=0.8\columnwidth]{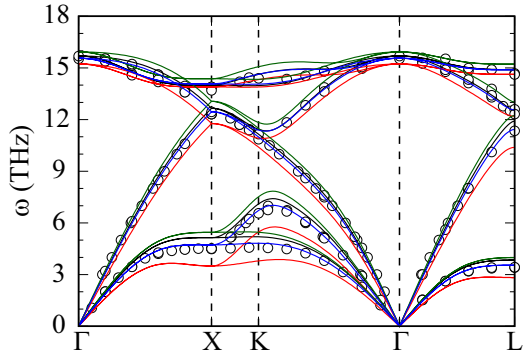}
	\end{center}
	\caption{
	The phonon bands of Si obtained at experimental lattice constant using NAO (black), LOAO (blue), the derivative of the inverse square root of the overlap matrix only (red), and the derivative of the atomic wave function only (dark-green). Open circles denote the measurement~\cite{76,77,81}.	}
	\label{Fig:Proj}
\end{figure}

We discuss the choice of atomic-site projectors for force calculation of group-IV semiconductors. The relaxed structures of Si with NAO and LOAO have minor differences: the full relaxed lattice constants from the former and latter projectors are 5.439 and 5.434, respectively. When the overestimation of the occupation number is not crucial, the structure optimization with NAO is expected to have marginal errors ~\cite{89}. However, the phonon dispersions may be affected by the small change in forces. 

Fig.~\ref{Fig:Proj} shows that the phonon dispersion of Si at the experimental lattice constant. The Pulay forces using NAO basis overestimate the phonon dispersion in the acoustic modes compared to measurement at ambient conditions. On the other hand, the Pulay forces ($F^N$ in Eq.~\ref{Eq:Pulay}) using LOAO basis give agreeable phonon dispersion for entire modes. Moreover, the derivate of the inverse square root of the overlap matrix has significant effect on the Pulay forces as previously reported~\cite{39}. Without it, the overall phonon dispersion is overestimated compared to experiment. Both the derivate of the inverse square root of the overlap matrix and the derivate of the atomic wave functions should be included in the Pulay forces in the LOAO basis.

\bibliography{eACBN0_force}

\end{document}